\documentclass[12pt]{article}
\usepackage{graphicx}
\usepackage[hang,small,bf]{caption}
\newcommand{\be}[3]{\begin{equation}  \label{#1#2#3}}

\newcommand{\ee}{ \end{equation}}
\newcommand{\ba}{\begin{array}}
\newcommand{\ea}{\end{array}}

\let\LARGE=\Large
\let\Large=\large

\begin{document}


\thispagestyle{empty}
\rightline{UPR-868-T}
\rightline{CALT-68-2259}
\rightline{CITUSC/00-007}
\rightline{hep-th/0001159}

\vspace{1truecm}

\centerline{\bf \LARGE 
Anti-deSitter Vacua of Gauged
}
\bigskip
\centerline{\bf \LARGE
Supergravities with 8 Supercharges}
\bigskip

\vspace{1.2truecm}
\centerline{\bf Klaus Behrndt$^a$\footnote{e-mail: 
 behrndt@theory.caltech.edu}
{\rm and} Mirjam Cveti{\v c}$^b$\footnote{e-mail: cvetic@cvetic.hep.upenn.edu}}
\vspace{.5truecm}
\centerline{$^a$ \em California Institute of Technology}
\centerline{\em Pasadena, CA 91125}

\bigskip

\centerline{\it CIT-USC Center For Theoretical Physics}
\centerline{\it University of Southern California}
\centerline{\it Los Angeles, CA 90089-2536}

\vspace{.3truecm}

\centerline{$^b$ \em Department of Physics and Astronomy}
\centerline{\em University of Pennsylvania, Philadelphia, PA 19104-6396}

\vspace{1truecm}


\begin{abstract}
We investigate supersymmetric extrema of Abelian gauged supergravity
theories with non-trivial vector multiplets and 8 supercharges in four
and five dimensions. The scalar fields of these models parameterize a
manifold consisting of disconnected branches and restricting to the case
where this manifold has a non-singular metric we show that on every
branch there can be at most one extremum, which is a local maximum 
(for $W>0$) or a
minimum  (for $W<0$)
of the superpotential $W$. Therefore, these supergravity models do
not allow for regular domain wall solutions interpolating between
different extrema of the superpotential and the space-time transverse to
the wall asymptotically always approaches the boundary of AdS (UV-fixed
points in a dual field theory).
\end{abstract}


\newpage


There has been renewed interest in supergravity theories which allow
for anti de Sitter (AdS) vacuum solutions. On one hand, the AdS/CFT
correspondence implies that domain wall solutions encode the
information on the renormalization group (RG) flow of (strongly
coupled) super Yang Mills theories as discussed in \cite{040}.  On the
other hand, in the Randall-Sundrum scenario \cite{060, 070} one
considers a domain wall in a five-dimensional AdS space-time which
allows for localization of gravity near the wall.  In both cases, the
gravitational effects in the domain wall backgrounds play an essential
role. In the context of fundamental theory it is essential to consider
supergravity theories with a non-trivial potential and demonstrate the
existence of the domain wall solutions with the desired gravitational
effects. The first examples of the superymmetric domain walls were
found in N=1 D=4 supergravity theory \cite{030} (for a review see
\cite {210}).  These solutions are static and interpolate between
isolated supersymmetric extrema of the scalar potential.  The
gravitational properties of these domain walls crucially depend on the
features of the superpotential and have been classified in
\cite{250,010}.

Unlike supergravity theories with four supercharges, e.g., N=1 D=4
supergravity, which have a rich structure of possible domain walls,
the field theory embedding of possible domain wall solutions becomes
highly constrained in supergravity theories with at least 8
supercharges, i.e.\ (N=1, D=5), (N=2, D=4) or (N=4, D=3)
supergravity. In these cases the structure of the potential is related
to gaugings of isometries of the scalar field manifold and thus it is
much more restricted. There are the following different possibilities
to gauge these supergravity models: (i) to gauge a subgroup of the
$SU(2)$-R-symmetry, (ii) to gauge isometries of the vector moduli
space or (iii) to gauge isometries of hyper-multiplet moduli
space. In four dimensions the different cases have been reviewed in
\cite{180} and the 5-d cases are discussed in \cite{110, 090, 220}.

We will show that for the case (i), i.e. for {\it Abelian gauged}
supergravity with eight supercharges, all extrema of the
superpotential are disconnected as long as we restrict ourselves to
scalar field manifolds with a non-singular metric. It is therefore
impossible to construct regular domain wall solutions interpolating
between different extrema. In addition, the space-time transverse to
the wall always approaches the boundary of AdS asymptotically and thus
these solutions disallow for the localization of gravity near the
interior of the wall, i.e. in D=5 it is impossible to embed the
Randall-Sundrum scenario in this framework.~\footnote{The same
conclusion seems to hold also for the case (ii)\cite{140,092}. On the
other hand, we restrict ourselves to studying vector multiplets
only and will not consider case (iii).}

{\bf 1) $D=5$ Case} For this case equivalent conclusions have been
derived in \cite{140}. In D=5 supergravity with 8 supercharges the
(real) scalars in the vector multiplets $\phi^A$ parameterize a
hypersurface ${\cal M}$ defined by a cubic equation \cite{120}:
\be010
F(X) \equiv {1 \over 6} \, C_{IJK} X^I X^J X^K = 1
\ee
where $I = 0,1,2, \ldots n$ and $n$ is the number of vector multiplets
and $C_{IJK}$ are the coefficient defining the cubic Chern-Simons term in
the supergravity Lagrangian. (In Calabi-Yau compactifications these are
the topological intersection numbers.) In general ${\cal M}$ is not
connected and consists of different branches, separated by regions where
$F(X) < 0$; see figure 1.

\begin{figure}  \label{fig1}
\begin{center}
\includegraphics[angle = -90,width=60mm]{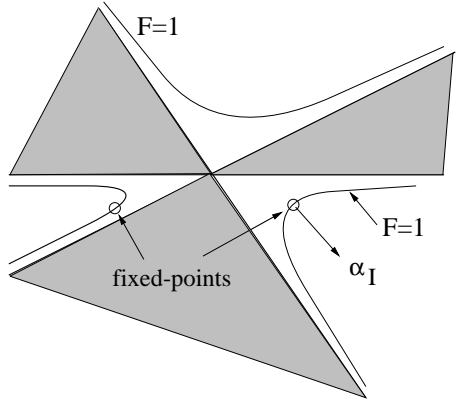}
\end{center}
\caption{The scalar fields of vector super-multiplets of 
D=5 theory parameterize a
manifold that consists of different branches and due to the attractor
equations point where the normal vector is parallel to a given constant
vector $\alpha_I$ are ``fixed-points'' or extrema of the superpotential.
The straight lines correspond to F=0 domain and shaded areas to $F < 0$
domains.}
\end{figure}

Gauging a $U(1)$ subgroup of the $SU(2)$ R-symmetry \cite{110}, the
scalars remain uncharged, however they  obtain a potential given by
\be100
V = 6 \Big( \, {3 \over 4} g^{AB} \partial_A W \partial_B W - W^2 \, \Big)\ , 
\ee
with the superpotential and the metric:
\be110
W = \alpha_I X^I \ , \ \ g_{AB}= - {1 \over 2}  \Big(\partial_A X^I 
\partial_B X^J\, \partial_I \partial_J F(X)\Big)\Big|_{F=1}  \ ,
\ee
{From} the M/string-theory perspective this superpotential appears due
to calibrated sub-manifolds of the internal space, where the vector
$\alpha_I$ corresponds to non-trivial fluxes (see \cite{220,200}). On the
other hand, from D=5 perspective this superpotential arises solely due
to constraints of supersymmetry \cite{080, 110}.  Supersymmetric
extrema of $V$ are given by extrema of $W$ and because $\partial_A
X^I$ defines tangent vectors on ${\cal M}$, supersymmetric extrema are
points on ${\cal M}$ where the constant vector $\alpha_I$ is normal to
${\cal M}$, i.e.\ where
\be120
\alpha_I \sim X_I,
\ee
with $X_I = {1\over 3} \partial_I F(X)\big|_{F=1}$ (see also figure
1). This is the essence of the attractor equation as derived in
\cite{150, 160}, which has been discussed in the domain wall context
in \cite{020, 260}.

In addition,  the second derivative of $W$ satisfies the following constraint
\cite{120}:
\be130
\partial_A \partial_B W = {2\over 3} g_{AB} W + T_{ABC}g^{CE}\partial_E W\ ,
\ee
where $T_{ABC} \sim \partial_A X^I \partial_B X^J \partial_C X^K C_{IJK}$.

At the extrema of the superpotential (fixed-points) $\partial_E W=0$ and
thus eq. (\ref{130}) implies that for a positive definite scalar metric
$g_{AB}$ these extrema can only be {\it minima } (for $W>0$) or {\it maxima}
(for $W<0$), but {\it not saddle points}. Moreover this equation implies that
the supersymmetric extrema always correspond to the {\it maxima} of the
potential, i.e., $\partial_{A}\partial_B V= - 4 g_{AB} W^2$,
see also \cite{262}.

Such two extrema, (see figure 2) could be connected if one allows for a
saddle point in-between, however since we restricted ourselves to the
physical domain of the scalar metric, i.e. we assume that $g_{AB}$ is
positive definite, such saddle points are excluded and all the extrema of
$W$ lie on disconnected branches of ${\cal M}$.~\footnote{ One arrives at
the same conclusion if one assumes that ${\cal M}$ is convex, which
implies that there can only be one point on any given branch where
(\ref{120}) holds \cite{170}.}

\begin{figure}
\begin{center}
\includegraphics[angle = -90,width=70mm]{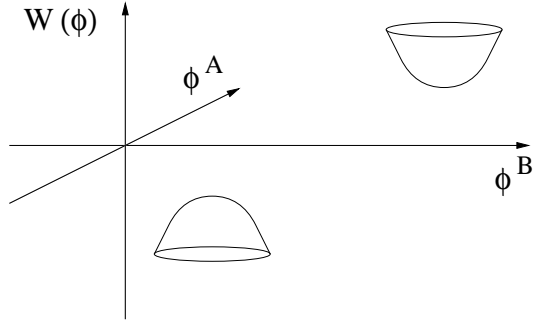}
\end{center}
\caption{Restricting to a region where the kinetic part of the scalar
fields is strictly positive definite, any critical point, where
$\partial_A W(\phi^A) =0$, is  either a minimum (for $W>0$)  or a maximum (for
$W<0$). Since  saddle points are also excluded, the critical
points cannot be connected in a regular way.
}
\end{figure}

Let us also point out that the space-time transverse to the wall
necessarily asymptotes to the boundary of AdS and not the Cauchy AdS
horizon, thus exceeding the one which is necessary for implementation of
the Randall-Sundrum set-up. (Equivalent arguments are given in
\cite{140}.) Namely, for the static domain wall Ansatz:
\be111
ds^2= {\cal A}(z) (-dt^2+\sum dx_i^2) +dz^2
\ee
the Killing spinor equations which fixes the scalars $\phi^A(z)$
take the form 
\cite{010,230,232}:
\be222
\partial_z \phi^A=\pm 3 g^{AB} \partial_B W
\quad , \quad 
\partial_z \log {\cal A} = \mp 2 W,
\ee
with the spinor constraint $\Gamma_z \epsilon = \pm \epsilon$. The
expansion of the kink solution $\phi^A=\phi^A_{|\pm}+\delta
\phi^A$  around the supersymmetric extremum ($\partial _B
W_{|\pm}=0$) renders (\ref{222}) in the following asymptotic form:
\be333
\partial_z (\log\delta\phi^A)= \pm 2W_{|\pm}\ , \ \ \partial_z(\log 
{\cal A})= \mp 2W _{|\pm}\ .
\ee
(In the derivation of the first eq. in (\ref{333}) we employed
(\ref{130}), evaluated at $\partial_BW=0$.) The first eq. in (\ref{333})
implies that for a kink solution to approach (exponentially fast) the
asymptotic values $\phi^A_{\pm }$ (as $z\to \pm \infty$) the
superpotential $W$ has to satisfy: $\rm{ sign} W_+=-\rm{sign}W_-$.  As a
consequence, the second eq. in (\ref{333}) implies that in this case the
metric coefficient $\cal A$ necessarily grows exponentially fast on
either side of the wall. Thus, these walls, in addition to being
singular, necessarily approach the boundary of the AdS space-time as
$z\to \pm \infty$; they have a {\it repulsive gravity on either side} of
the wall and thus {\it cannot} localize gravity. In a dual field theory
these supersymmetric extrema always correspond to ultra-violet (UV)
fixed-points \cite{140}. The one-scalar example of such walls were given in
\cite{010,260} (see also \cite{280} for an early work on supergravity
kinks); in the interior these walls have a
power-law curvature singularity.

Another comment is in order.  If one does not insist on the positive
definite scalar metric $g_{AB}$, some supersymmetric extrema can
become saddle points\footnote{For special parameter choices the
extrema of $W$ may be at the boundary of ${\cal M}$ and may not
correspond to AdS vacua. We thank S.\ Gubser for communications on
this point.}. In this case it is possible to connect, e.g., a
supersymmetric maximum with a supersymmetric saddle point in a
continuous manner. However, again due to (\ref{333}), the space-time on
either side of such non-singular walls asymptotes to the boundaries of
AdS, which are UV fixed points of the dual field theories.

\bigskip

{\bf 2) $D=4$ Case} As the second example we consider D=4 , \ N=2 gauged
supergravity (for a review see \cite{180}).  In contrast to the  D=5 case
before, the scalars of vector supermultiplets  are now complex and the potential is given by
\be140
V =e^{K} \Big(\, g^{A\bar B} D_A W D_{\bar B} {\bar{W}} - 3 |W|^2 \, \Big),
\ee
where $W$ is the superpotential, $K$ is the K\"ahler potential, $D_A W
\equiv \Big(\partial_A + (\partial_A K)\Big)W$, and $g_{A \bar
B}=\partial_A \partial_{\bar B} K $ is the K\"ahler metric. In
comparison with 5-d supergravity we have to replace the scalars $X^I$
by the symplectic section $(X^I , F_I)$ where $F_I = \partial_I F(X)$
denotes is derivative of the prepotential $F(X)$. The superpotential
is again a linear function, but now in the symplectic section
\cite{190, 200, 290}:
\be150
W = \alpha_I X^I - \beta^I F_I \ .
\ee
Supersymmetric extrema of $V$ are given by extrema of $W$ with respect to
the {\it covarient} derivatives, i.e.\  $D_A W =0$. 
In order to facilitate the investigation of supersymmetric
extrema we write the potential in terms of a real function $\widehat W$:
\be160
\widehat W \equiv \xi 
|W |e^{K/2} = \xi |\alpha_I L^I - \beta^I M_I| \ ,
\ee
which is invariant under K\"ahler transformations and the analogous
constraint to (\ref{010}) becomes $i(\bar L^I M_I - L^I \bar M_I) =1$.
Here $\xi =\pm 1$ and can only change sign iff $W$ passes through zero.

In terms of $\widehat W$, the potential (\ref{140}) takes the form:
\be180
V = 3 \Big( \, {4 \over 3} g^{A \bar B} 
\partial_A \widehat W \partial_{\bar B}
\widehat W - \widehat W^2 \, \Big).
\ee 
Since $\widehat W$ satisfies the relation: ${(\partial_A \widehat W)
{\widehat W}^{-1}} = {(D_A W)}(2 W)^{-1}$, extrema of $\widehat W$
correspond to the supersymmetric extrema of the potential. Note, that
employing the real function $\widehat W$ the potential (\ref{180}) has
been cast in a form completely parallel to that of D=5 potential.  In
order to obtain the second derivatives at extrema, we can employ basic
formulae from special geometry.  Namely, the symplectic section ${\cal
V} = (L^I , M_I) = e^{K/2} (X^I , F_I)$ satisfies \cite{180}
\be182
D_A D_B {\cal V} = i C_{ABC} g^{C\bar E} D_{\bar E} \bar {\cal V}
\quad , \quad D_A D_{\bar B} {\cal V} = g_{A \bar B} {\cal V} \ ,
\ee
where $C_{ABC}$ is the covariantly holomorphic section. Moreover,
using the definition of $\widehat W$ one finds for the second
derivatives: ${(\partial_A \partial_B \widehat W) {\widehat W}^{-1}} =
{(D_A D_B W) (2 W)^{-1}} + {\cal O}(D_A W)$ and because $D_A {\cal V}
\equiv (\partial_A + {1 \over 2} (\partial_A K)){\cal V} = e^{K/2} D_A
W$ we obtain ${(\partial_A \partial_B \widehat W) {\widehat W}^{-1}} =
(D_A D_B {\cal V})(2{\cal V})^{-1} + {\cal O}(D_A W)$. As a
consequence of (\ref{182}) we find that at supersymmetric extrema
($\partial_A {\widehat W}=0$) the second derivatives of $\widehat W$
satisfy:
\be190
{\partial_A \partial_{\bar B} \widehat W}  = {1 \over 2}
g_{A \bar B} \widehat W \quad, \quad \partial_A \partial_B \widehat W = 0 \ .
\ee
The relationships (\ref{190}) imply the same conclusions as in D=5
case: for the domain with a positive definite K\"ahler metric
$g_{A{\bar B}}$, i.e.  restricting to the physical region of the
metric, all the extrema of $\widehat W$ are either minima (for
$\widehat W > 0$) or maxima (for $\widehat W < 0$), but never saddle
points. This result again implies that the supersymmetric extrema are
disconnected and that the potential always has maxima there,
i.e. $\partial_{A}\partial_{\bar B}V= -g_{A\bar B}{\widehat W}^2$ and
$\partial_A\partial_B V=0$.

For the purpose of addressing the space-time properties of
supersymmetric (static) domain wall backgrounds one arrives at the
following Killing spinor equations which we cast in an explicitly
K\"ahler invariant form:
\be322
\partial_z \phi^A=-2g^{A\bar B} \partial_{\bar B} {\widehat W}\ 
, \ \ 
\partial_z \log {\cal A} ={\widehat W}\ .
\ee
In addition,  the complex scalar fields have to statisfy the 
``geodesic equation'':
\be325
{\rm Im}\left[(\partial_z \phi^A) \partial_A (\log {\widehat W})\right]=0
\ ,
\ee
while the Killing spinors satisfy: $\epsilon_{\alpha} =\xi i \gamma^z
e^{\theta_W} \varepsilon_{\alpha\beta} \epsilon^{\beta}$,
where $\theta_W$ is the phase of the holomorphic superpotential $W$. 
The geodesic eq. (\ref{325}) is a supergravity generalization of the
equation in global supersymmetric theory where a kink solution
corresponds to a straight line in the W-plane ($\partial_z\theta_W=0$)
(see, e.g., \cite{270,240}). The above Killing spinor equations were
first derived for domain walls in D=4 N=1 supergravity \cite {030} ,
where the superpotential $W$ and K\"ahler potential $K$ are not
subject to constraints of N=2 special geometry; of course for the N=1
case the equations remain the same, but with constrained $K$ and
$W$.

The expansion of the kink solution $\phi^A=\phi^A_{|\pm}+\delta
\phi^A$  around the supersymmetric extremum ($\partial _B
{\widehat W}_{|\pm}=0$) renders (\ref{322}) in the following
asymptotic form:
\be400
\partial_z (\log\delta\phi^A)=-2{\widehat W}_{\pm}\ , \ 
\ \partial_z(\log {\cal A})= 2{\widehat W}_{\pm} \ .
\ee
In the derivation of the first eq. in (\ref{400}) we used the
relationships (\ref{190}). The first eq. in (\ref{400}) implies that
that for the existence of a kink solution ($\phi^A \to \phi^A_{|\pm}$
as $z\to \pm \infty$) the superpotential $W$ necessarily crosses zero
and thus $\xi_{|+}=-\xi_{|-}$. The second eq. in (\ref{400}) in turn
implies that in this case the metric coefficient $\cal A$ necessarily
grows exponentially fast on either side of the wall.  Thus, just as in
the D=5 case, these walls are necessarily singular (because extrema
are on different branches) and the space-time asymptotically
approaching the boundary of the AdS on either side of the wall (UV
fixed points). On the other hand, just as in the D=5 case, if one
relaxes the constraint of positive definite K\"ahler metric, such
domains could connect across a smooth region, but the asymptotic
space-time remains to approach the boundary of AdS asymptotically.

\bigskip 

We have not considered the D=3 case with 8 supercharges, where the
corresponding scalars parameterize a quarternionic manifold; we
expect, however, the same conclusions. On the other hand, just as for
the D=4 cases, breaking further supersymmetry (to four or only two
supercharges) one expects a much richer structure (see \cite{400}).

\bigskip

Let us end with some general remarks. In our arguments it was
important to assume that the K\"ahler metric is everywhere positive
definite, which excluded saddle points and disconnected all
supersymmetric extrema.  This is a very strong restriction, which may
not be the case for physically interesting applications. E.g., the
manifold ${\cal M}$ can have boundaries where eigenvalues of the
K\"ahler metric vanish and additional massless modes are expected.  In
addition, we restricted ourselves to supersymmtric cases only, but it
may be that the supersymmetric vacua are connected by non-BPS
sphaleron configurations as recently discussed in \cite{420}.  These
are very interesting aspects, which certainly deserve further
investigations.

\bigskip


\bigskip

{\bf Acknowledgments}

The work is supported by a DFG Heisenberg grant (K.B.), in part by the
Department of Energy under grant number DE-FG03-92-ER 40701 (K.B.),
DOE-FG02-95ER40893 (M.C.) and the University of Pennsylvania Research
Foundation (M.C). M.C. would like to thank Caltech High Energy Theory
Group for hospitality during the completion of the work.


%
%

\begin{thebibliography}{10}

\bibitem{040}
D.~Z. Freedman, S.~S. Gubser, K.~Pilch, and N.~P. Warner, ``Renormalization
  group flows from holography supersymmetry and a c-theorem,''
  \href{http://xxx.lanl.gov/abs/hep-th/9904017}{{\tt hep-th/9904017}}.

\bibitem{060}
L.~Randall and R.~Sundrum, ``An alternative to compactification,'' {\em Phys.
  Rev. Lett.} {\bf 83} (1999) 4690,
  \href{http://xxx.lanl.gov/abs/hep-th/9906064}{{\tt hep-th/9906064}}.

\bibitem{070}
L.~Randall and R.~Sundrum, ``A large mass hierarchy from a small extra
  dimension,'' {\em Phys. Rev. Lett.} {\bf 83} (1999) 3370--3373,
  \href{http://xxx.lanl.gov/abs/hep-ph/9905221}{{\tt hep-ph/9905221}}.

\bibitem{030}
M.~Cveti{\v c}, S.~Griffies, and S.-J. Rey, ``Static domain walls in N=1
  supergravity,'' {\em Nucl. Phys.} {\bf B381} (1992) 301--328,
  \href{http://xxx.lanl.gov/abs/hep-th/9201007}{{\tt hep-th/9201007}}.

\bibitem{210}
M.~Cveti{\v c} and H.~H. Soleng, ``Supergravity domain walls,'' 
{\em Phys. Rept.}
  {\bf 282} (1997) 159, \href{http://xxx.lanl.gov/abs/hep-th/9604090}{{\tt
  hep-th/9604090}}.

\bibitem{250}
M.~Cveti{\v c} and S.~Griffies, 
``Gravitational effects in supersymmetric domain
  wall backgrounds,'' {\em Phys. Lett.} {\bf B285} (1992) 27--34,
  \href{http://xxx.lanl.gov/abs/hep-th/9204031}{{\tt hep-th/9204031}}.

\bibitem{010}
K.~Behrndt and M.~Cveti{\v c}, ``Supersymmetric domain wall world 
from {D=5} simple
  gauged supergravity,'' \href{http://xxx.lanl.gov/abs/hep-th/9909058}{{\tt
  hep-th/9909058}}.

\bibitem{180}
L.~Andrianopoli {\em et.~al.}, ``General matter coupled {N=2} supergravity,''
  {\em Nucl. Phys.} {\bf B476} (1996) 397--417,
  \href{http://xxx.lanl.gov/abs/hep-th/9603004}{{\tt hep-th/9603004}}.

\bibitem{110}
M.~Gunaydin, G.~Sierra, and P.~K. Townsend, ``Gauging the {D=5}
  {M}axwell-{E}instein supergravity theories: More on {J}ordan algebras,'' {\em
  Nucl. Phys.} {\bf B253} (1985) 573.

\bibitem{090}
M.~Gunaydin and M.~Zagermann, ``The gauging of five-dimensional, {N} = 2
  {Maxwell-Einstein} supergravity theories coupled to tensor multiplets,''
  \href{http://xxx.lanl.gov/abs/hep-th/9912027}{{\tt hep-th/9912027}}.

\bibitem{220}
A.~Lukas, B.~A. Ovrut, K.~S. Stelle, and D.~Waldram, ``Heterotic {M}-theory in
  five dimensions,'' {\em Phys. Rev.} {\bf D59} (1999) 086001,
  \href{http://xxx.lanl.gov/abs/hep-th/9806051}{{\tt hep-th/9806051}}.

\bibitem{140}
R.~Kallosh and A.~Linde, ``Supersymmetry and the brane world,''
  \href{http://xxx.lanl.gov/abs/hep-th/0001071}{{\tt hep-th/0001071}}.

\bibitem{092}
M.~Gunaydin and M.~Zagermann, unpublished.

\bibitem{120}
M.~Gunaydin, G.~Sierra, and P.~K. Townsend, ``The geometry of {N=2}
  {M}axwell-{E}instein supergravity and {J}ordan algebras,'' {\em Nucl. Phys.}
  {\bf B242} (1984) 244.

\bibitem{200}
K.~Behrndt and S.~Gukov, ``Domain walls and superpotentials from {M} theory on
  {Calabi-Yau} three-folds,''
  \href{http://xxx.lanl.gov/abs/hep-th/0001082}{{\tt hep-th/0001082}}.

\bibitem{080}
P.~K. Townsend, ``Positive energy and the scalar potential in higher
  dimensional (super)gravity theories,'' {\em Phys. Lett.} {\bf B148} (1984)
  55.

\bibitem{150}
S.~Ferrara and R.~Kallosh, ``Universality of supersymmetric attractors,'' {\em
  Phys. Rev.} {\bf D54} (1996) 1525--1534,
  \href{http://xxx.lanl.gov/abs/hep-th/9603090}{{\tt hep-th/9603090}}.

\bibitem{160}
S.~Ferrara and R.~Kallosh, ``Supersymmetry and attractors,'' {\em Phys. Rev.}
  {\bf D54} (1996) 1514--1524,
  \href{http://xxx.lanl.gov/abs/hep-th/9602136}{{\tt hep-th/9602136}}.

\bibitem{020}
K.~Behrndt, ``Domain walls of {D} = 5 supergravity and fixpoints of {N} = 1
  super yang-mills,'' \href{http://xxx.lanl.gov/abs/hep-th/9907070}{{\tt
  hep-th/9907070}}.

\bibitem{260}
R.~Kallosh, A.~Linde, and M.~Shmakova, ``Supersymmetric multiple basin
  attractors,'' {\em JHEP} {\bf 11} (1999) 010,
  \href{http://xxx.lanl.gov/abs/hep-th/9910021}{{\tt hep-th/9910021}}.

\bibitem{262} 
Aaron Chou and others, ``Critical points and phase transitions in 5d
                  compactifications of  M-theory'', {\em Nucl. Phys.}
{\bf B508} (1997) 147--180, {\tt hep-th/9704142}.

\bibitem{170}
M.~Wijnholt and S.~Zhukov, ``On the uniqueness of black hole attractors,''
  \href{http://xxx.lanl.gov/abs/hep-th/9912002}{{\tt hep-th/9912002}}.

\bibitem{230}
D.~Z. Freedman, S.~S. Gubser, K.~Pilch, and N.~P. Warner, ``Continuous
  distributions of {D3}-branes and gauged supergravity,''
  \href{http://xxx.lanl.gov/abs/hep-th/9906194}{{\tt hep-th/9906194}}.

\bibitem{232}
K.~Skenderis and P.~K. Townsend, ``Gravitational stability and
  renormalization-group flow,'' {\em Phys. Lett.} {\bf B468} (1999) 46,
  \href{http://xxx.lanl.gov/abs/hep-th/9909070}{{\tt hep-th/9909070}}.

\bibitem{280}
M.~Gunaydin, G.~Sierra, and P.~K. Townsend, ``More on {D=5}
  {M}axwell-{E}instein supergravity: Symmetric spaces and kinks,'' {\em Class.
  Quant. Grav.} {\bf 3} (1986) 763.

\bibitem{190}
T.~R. Taylor and C.~Vafa, ``{RR} flux on {Calabi-Yau} and partial supersymmetry
  breaking,'' \href{http://xxx.lanl.gov/abs/hep-th/9912152}{{\tt
  hep-th/9912152}}.

\bibitem{290}
J.~Polchinski and A.~Strominger, ``New vacua for type {II} string theory,''
  {\em Phys. Lett.} {\bf B388} (1996) 736--742,
  \href{http://xxx.lanl.gov/abs/hep-th/9510227}{{\tt hep-th/9510227}}.

\bibitem{270}
P.~Fendley, S.~D. Mathur, C.~Vafa, and N.~P. Warner, ``Integrable deformations
  and scattering matrices for the {N=2} supersymmetric discrete series,'' {\em
  Phys. Lett.} {\bf B243} (1990) 257--264.

\bibitem{240}
M.~Cveti{\v c}, F.~Quevedo, and S.-J. Rey, 
``Stringy domain walls and target space
  modular invariance,'' {\em Phys. Rev. Lett.} {\bf 67} (1991) 1836--1839.

\bibitem{400}
N.S.~Deger and A.~Kaya and E.~Sezgin and P.~Sundell,
``Matter coupled $AdS_3$ supergravities and their black strings'',
{\tt hep-th/9908089}.

\bibitem{420}
J.A.~Harvey, P.~Horava and P.~Kraus, ``D-sphalerons and the
toplogy of string configuration space'', {\tt hep-th/0001143}.


\end{thebibliography}

\providecommand{\href}[2]{#2}\begingroup\raggedright\endgroup


\end{document}